\documentclass[12pt]{iopart}
\usepackage{iopams}
\usepackage[dvips]{graphicx}
\newcommand{\VEV}[1]{\langle{#1}\rangle}
\newcommand{\chibar}{{\bar{\chi}}}
\newcommand{\Psfig}[3]{\includegraphics[width=#1 #3]{QM06-Ohnishi-Figs/#2}}
\begin{document}

\title[Phase diagram in the strong coupling region of lattice QCD]
{Phase diagram at finite temperature and quark density
in the strong coupling region of lattice QCD for color SU(3)}

\author{A Ohnishi, N Kawamoto, K Miura}

\address{Department of Physics, Faculty of Science, Hokkaido University\\
Sapporo 060-0810, Japan}
\ead{ohnishi@nucl.sci.hokudai.ac.jp}
\begin{abstract}
We study the phase diagram of quark matter
at finite temperature ($T$) and chemical potential ($\mu$)
in the strong coupling region of lattice QCD for color SU(3).
Baryon has effects to extend the hadron phase to a larger $\mu$ direction
relative to $T_c$ at low temperatures in the strong coupling limit.
With the $1/g^2$ corrections,
$T_c$ is found to decrease rapidly as $g$ decreases,
and the shape of the phase diagram becomes closer
to that expected in the real world.
\end{abstract}


\section{Introduction}
\label{Sec:Introduction}

Exploring various phases of quark and nuclear matter
is one of the primary goals in high-energy heavy-ion collision physics.
Compared to the QCD phase transition at high $T$,
less is known for the cold and dense matter,
partly because the lattice Monte-Carlo simulations are difficult
at large $\mu$.
One of the most instructive
approaches to compressed baryonic matter is to consider
the strong coupling limit (SCL) of lattice 
QCD~\cite{Kawamoto,Faldt1986,Bilic1992,Nishida2004}.
In fact, effective free energies at finite $T$ and $\mu$
have been analytically derived in SCL,
and it is successfully applied to nuclear many-body problems~\cite{Tsubakihara}.
While baryon effects would be important in dense matter,
they have been ignored in finite $T$ treatments
through the $1/d$ expansion ($d$ is the spatial dimension).

In this proceedings, first we derive an expression of the effective free energy
at finite $T$ and $\mu$ including baryon effects~\cite{KMOO_2007}.
We find the second and first order phase transition at high and low $T$,
separated by the tricritical point (TCP) in the chiral limit. 
With finite quark masses,
these second order phase boundary and TCP
become the cross over and the critical end point,
thus the obtained phase diagram seems to show essential qualitative features
of that in the real world.

One of the problems in SCL
is in the ratio, $R_{\mu T} = \mu_c(T=0)/T_c(\mu=0)$;
this ratio would be larger than two in the real world,
but it is less than one half in all of the works based on SCL.
In the second part of this proceedings, 
we demonstrate that
the ratio increases to $R_{\mu T} \sim 1.8$ at $g \sim 1$
with $1/g^2$ corrections.
With finite bare quark mass and baryon effects,
it may be possible to understand the shape of the actual phase diagram.

\section{Effective free energy in the strong coupling limit of lattice QCD}
\label{Sec:Model}

In SCL, we ignore pure gluonic action ($\propto 1/g^2$),
and we obtain the following staggered fermion effective action
after integrating spatial links,
\begin{equation}
{\cal Z}=\int {\cal D}[\chi,\chibar,U_0]
\exp\left[-S_F^{(0)}
	+\frac12 (M, V_M M)
	+(\bar{B},V_B B)
\right]\,,
\label{Eq:ActionA}
\end{equation}
where
$(A,B)=\sum_x A_xB_x$,
and 
the mesonic and baryonic composites and their propagators are
$M_x=\chibar^a_x\chi^a_x$, 
$B_x=\varepsilon_{abc}\chi^a_x\chi^b_x\chi^c_x/6$,
$V_M(x,y)$
and
$V_B(x,y)$.
We keep the timelike link and bare quark mass terms
unexpanded,
$S_F^{(0)}=\sum_{x}[\chibar_xe^{\mu}U_0(x)\chi_{x+\hat{0}}
	-\chibar_{x+\hat{0}}e^{-\mu}U^\dagger_0(x)\chi_x]/2
	+m_0(\chibar,\chi)$.

We decompose the effective action in Eq.~(\ref{Eq:ActionA}),
containing six fermion terms,
to the bilinear form in $\chi$
in order to perform the quark integral in a finite $T$ treatment.
Decomposition has been carried out in three steps,
\numparts
\begin{eqnarray}
&&e^{(\bar{B},V_B,B)}
=\int {\cal D}[\bar{b},b,\phi,\phi^\dagger]
	e^{
		-(\bar{b},V_B^{-1}b)-\phi_a^\dagger\phi_a
		+\phi_a^\dagger D_a+D^\dagger_a\phi_a
		- \gamma^2 M^2/2+M\bar{b}b/9\gamma^2
	}
	\ ,
\label{Eq:DecompA}
\\
&&e^{M \bar{b}b/9\gamma^2}
=\int d[\omega]\,e^{
		-\omega^2/2
		-\omega(\alpha M+\bar{b}b/9\alpha\gamma^2)
		-\alpha^2 M^2/2
		}
\ ,
\label{Eq:DecompB}
\\
&&e^{\frac12(M,\widetilde{V}_M M)}
=\int d[\sigma]\,e^{
		-(\sigma,\widetilde{V}_M\sigma)/2
		-(\widetilde{V}_M\sigma,M)
		}
\label{Eq:DecompC}
\ .
\end{eqnarray}
\endnumparts
In the first step (\ref{Eq:DecompA}),
the baryonic action is decomposed 
using the diquark field $\phi_a$~\cite{Azcoiti2003},
whose expectation value is $\VEV{\phi_a}=\VEV{D_a}$,
where $D_a=\gamma\varepsilon_{abc}\chi^b\chi^c/2+\bar{\chi}^a b/3\gamma$.
In the second step (\ref{Eq:DecompB}),
the four fermi interaction term, $M\bar{b}b$,
is decomposed utilizing the null potency, $(\bar{b}b)^2=0$.
In the third step (\ref{Eq:DecompC}),
chiral condensate $\sigma$ is introduced
after including the remaining four quark interaction terms
in the propagator,
$\widetilde{V}_M=V_M+(\alpha^2+\gamma^2)\delta_{x,y}$.
The effective action is now in a bilinear and pfaffian form of $\chi$,
and quarks are separated in each spatial point
coupling to the auxiliary fields,
$m_q=\widetilde{V}_M\sigma+\alpha\omega$,
\begin{eqnarray}
\fl~~~~~~~~~~~~
S_F&=&
S_F^{(0)}
+(\chibar,m_q\chi)
+\frac12(\sigma,\widetilde{V}_M\sigma)
+(\bar{b},\widetilde{V}_B^{-1} b)
+\frac12 (\omega,\omega)
\nonumber\\
\fl
&+&(\phi^\dagger,\phi)
+{1\over3\gamma}\left[
	 (\chibar^a, \phi_a^\dagger b)
	+(\bar{b}\phi_a,\chi^a)
	\right]
+\frac{\gamma}{2}\varepsilon_{cab}\left[
	 (\phi_c^\dagger, \chi^a \chi^b)
	+(\chibar^b\chibar^a,\phi_c)
	\right]
\,.
\label{Eq:ActionB}
\end{eqnarray}

We can analytically carry out
Matsubara frequency product of quarks and temporal link integral
in the mean field approximation
at zero diquark condensate.
At equilibrium, 
an approximate linear relation $\omega \propto \sigma$ holds,
and the effective free energy is found to be
\begin{eqnarray}
{\cal F}_\mathrm{eff}(\sigma)
&=&\frac12b_\sigma\sigma^2
+{\cal F}_\mathrm{eff}^{(q)}(b_\sigma\sigma)
+{\it \Delta}{\cal F}_\mathrm{eff}^{(b)}(g_\sigma\sigma)
\,,
\label{Eq:Feff}
\\
{\cal F}_\mathrm{eff}^{(q)}(m_q)
&=&-T\log\left[C_\sigma^3-\frac12C_\sigma
+\frac14\cosh\left(\frac{3\mu}{T}\right)\right]
\,,
\\
{\it \Delta}{\cal F}_\mathrm{eff}^{(b)}(m_b)
&\simeq&-f^{(b)}\left({\pi m_b\over8}\right)
\,, 
\end{eqnarray}
where $C_\sigma=\cosh(\mathrm{arcsinh}\,(m_q)/T)$
and 
$f^{(b)}(x)=\log(1+x^2)/2-[\arctan{x}-x+x^3/3]/x^3-3x^2/10$.
In this effective free energy ${\cal F}_\mathrm{eff}$,
we have two parameters, $b_\sigma$ and $g_\sigma$,
which are related to the decomposition parameters,
$\alpha$ and $\gamma$.
Without baryon effects,
the first two terms remain in the effective free energy (\ref{Eq:Feff})
with a fixed coefficient $b_\sigma=d/2N_c$~\cite{Nishida2004}.

\begin{figure}
\begin{minipage}{\textwidth}
\Psfig{8cm}{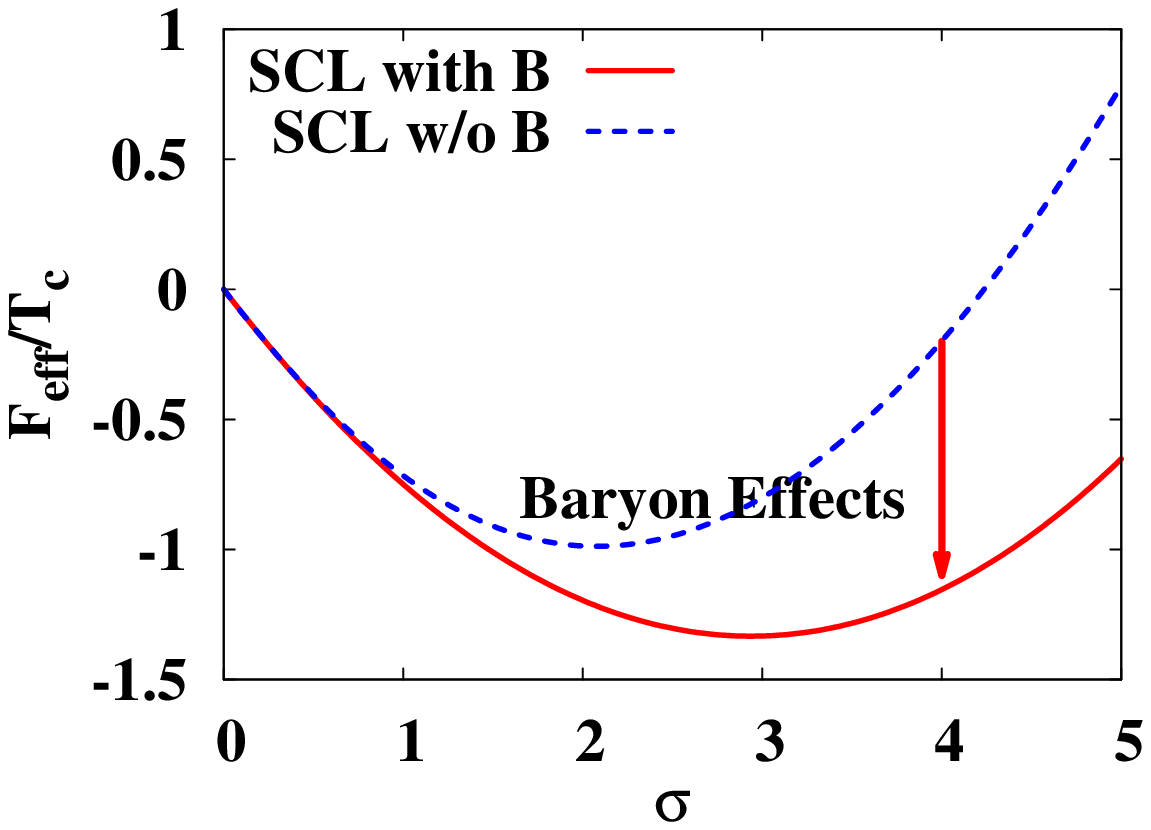}{}
\Psfig{8cm}{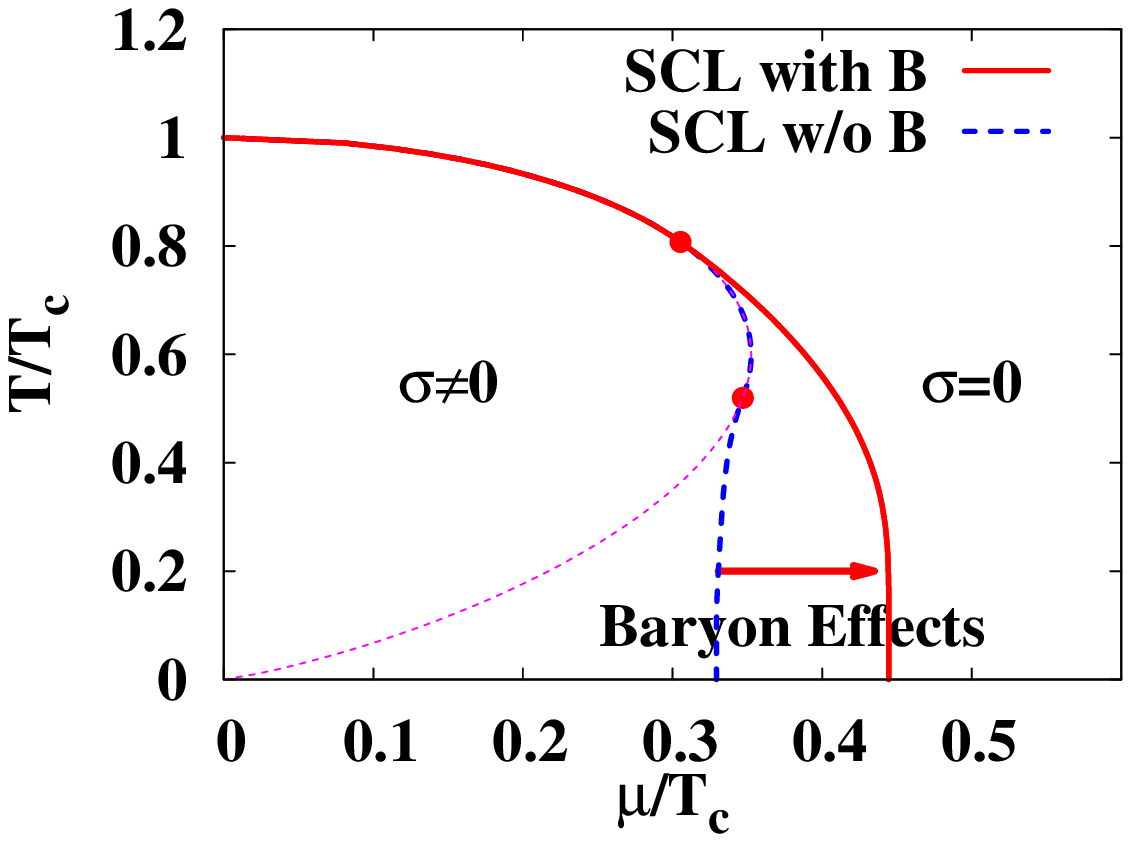}{}
\end{minipage}
\caption{Effective free energy as a function of $\sigma$ (left panel),
and the phase diagram in the strong coupling limit (right panel).
}\label{Fig:SCL}
\end{figure}

Baryon effects appear
in the modification of the coefficient $b_\sigma$
and in the additional term ${\it \Delta}{\cal F}_\mathrm{eff}^{(b)}$
coming from the auxiliary baryon determinant.
When we adopt the parameter $\alpha=0.2$
which approximately maximize the ratio
$R_{\mu T}=\mu_c/T_c$,
baryons are found to have effects
of the effective free energy gain with respect to $T_c$
and the extension of the hadronic phase in the larger $\mu$ direction,
as shown in Fig.~\ref{Fig:SCL}~\cite{KMOO_2007}.

\section{Finite coupling correction}
\label{Sec:Finite_g}

While the SCL seems to show
qualitative features of the phase diagram in the real world,
there are several problems,
such as
the parameter dependence and scale modifications~\cite{KMOO_2007},
no stable color superconductor (CSC) phase,
and too small ratio $R_{\mu T} < 0.5$.
Finite coupling corrections may solve
some of these problems~\cite{Faldt1986,Bilic1992,Bilic1995,Ipp}.

The $1/g^2$ correction on the effective action was
derived by Faldt and Petersson~\cite{Faldt1986},
and we can bosonize the plaquett contributions
in the mean field approximation as,
\begin{eqnarray}
\fl~~~~~~~~
{\it \Delta}S_F&=&
 	\frac{1}{4N_c^2g^2}\sum_{x,j>0} 
		(
		 V^\dagger_x V_{x+\hat{j}}
		+V^\dagger_x V_{x-\hat{j}}
		)
 	- \frac{1}{8N_c^4g^2}\sum_{x,k>j>0} 
		M_{x}
		M_{x+\hat{j}}
		M_{x+\hat{k}}
		M_{x+\hat{k}+\hat{j}}
\nonumber\\
\fl
&\simeq&
	N_s^3N_\tau\left(
		 \frac{\beta_t}{4}\varphi_t^2
		+\frac{\beta_sd}{4}\varphi_s^2
	\right)
	+\frac{\beta_t\varphi_t}{4}
	\sum_x(V_x-V_x^\dagger)
	-\beta_s\varphi_s
	\sum_{x,j>0} M_x M_{x+\hat{j}}
	\,.
\label{Eq:ActionC}
\end{eqnarray}
We have defined $V_x \equiv \chibar_x U_0(x) \chi_{x+\hat{0}}$,
$\beta_s=(d-1)/8N_c^4g^2$, and $\beta_t=d/2N_c^2g^2$,
and the auxiliary fields have expectation values of
$\VEV{\varphi_t}=\VEV{V^\dagger-V}$
and 
$\VEV{\varphi_s}=2\VEV{M_xM_{x+\hat{j}}}$.
Here we have omitted baryon effects (${\cal O}(1/d^{1/2})$)
and higher order terms (${\cal O}(1/d)$) 
in the $1/d$ expansion.

These corrections have a similar structure
to the SCL effective action (\ref{Eq:ActionA}),
and they lead to
the modifications of the quark mass and effective chemical potential as 
$\widetilde{m}_q=\sigma d(1+4N_c\beta_s\varphi_s-\beta_t\varphi_t\cosh\mu)/2N_c$
and
$\widetilde{\mu}=\mu-\beta_t\varphi_t\sinh\mu$.
At equilibrium, 
we can put $\varphi_s=2\sigma^2 + {\cal O}(1/g^2)$,
and the effective free energy up to ${\cal O}(1/g^2)$ is obtained as,
\begin{eqnarray}
{\cal F}_\mathrm{eff}=\frac{d}{4N_c}\sigma^2+3d\beta_s\sigma^4
	+\frac{\beta_t}{4}\varphi_t^2-N_c\beta_t\varphi_t\cosh\mu
	+{\cal F}_\mathrm{eff}^{(q)}(\widetilde{m}_q;T,\widetilde{\mu})
	\,.
	\label{Eq:FeffCorr}
\end{eqnarray}

\begin{figure}
\begin{minipage}{\textwidth}
\Psfig{6cm}{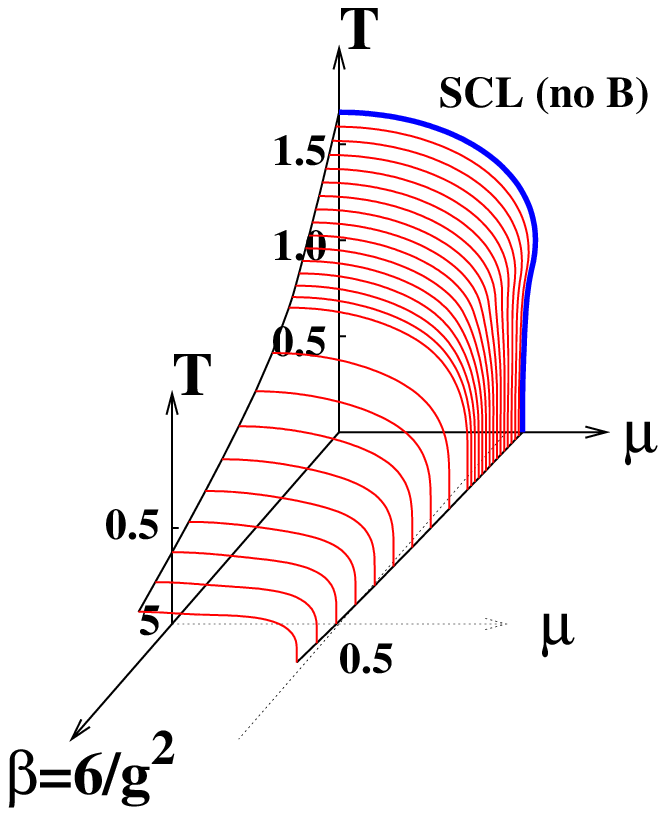}{}
\Psfig{8cm}{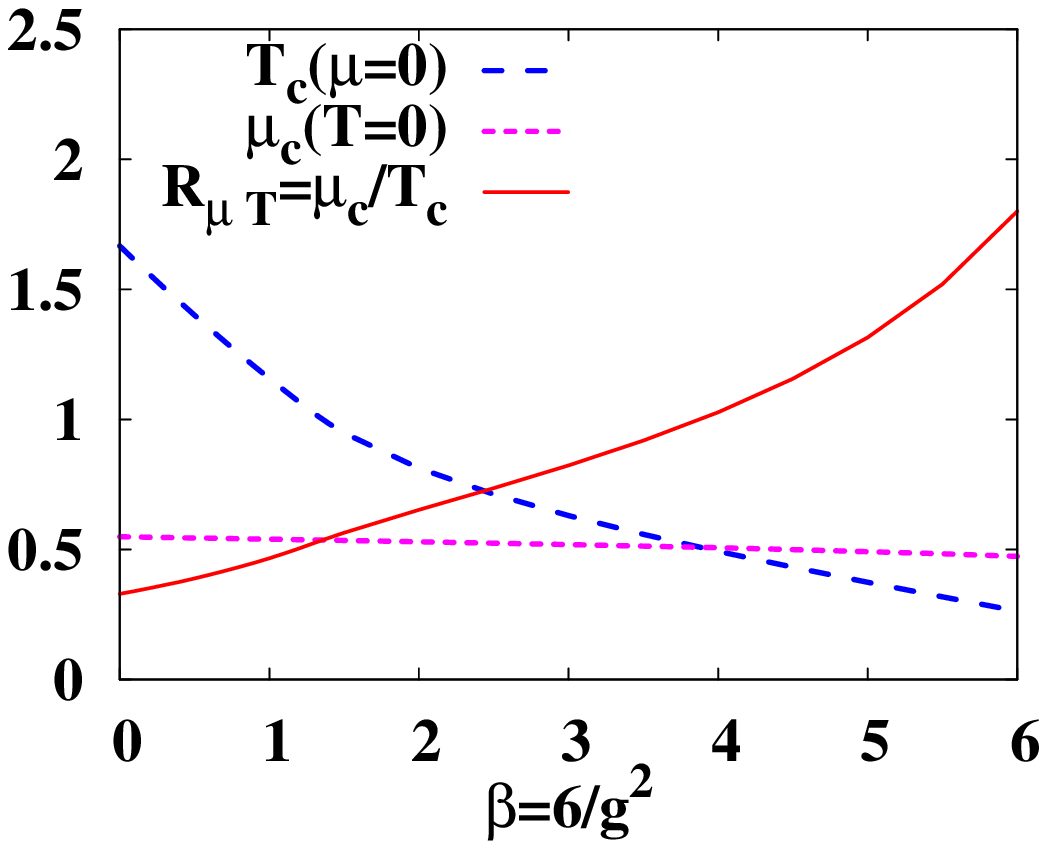}{}
\end{minipage}
\caption{Evolution of the phase diagram with $\beta=6/g^2$ (left panel),
and the $\beta$ dependence of $T_c$ and $\mu_c$ (right panel).}
\label{Fig:Finite_g}
\end{figure}

With this effective free energy,
$T_c$ is found to decrease as $g$ decreases,
while $\mu_c$ stays almost constant.
As a result, $R_{\mu T}$ grows to around 1.8 at $g \sim 1$
as shown in Fig.~\ref{Fig:Finite_g}.
The present results are not fully consistent with the previous 
findings ~\cite{Bilic1995}.
The difference in the bosonization scheme and lattice
anisotropy~\cite{Bilic1992,Bilic1995} has to be investigated further.

\section{Summary}

In this proceedings,
we have investigated the phase diagram
in the strong coupling region of lattice QCD.
Baryons are found to have effects of extending the hadron phase
in the larger $\mu$ direction with respect to $T_c$ up to around
30 \%~\cite{KMOO_2007} in the strong coupling limit.
With finite coupling corrections,
we have found that the ratio $R_{\mu T}=\mu_c/T_c$ becomes
closer to that expected in the real world.
It would be interesting to evaluate both of the finite coupling correction
and baryon effects simultaneously.

This work is supported in part by the Ministry of Education,
Science, Sports and Culture,
Grant-in-Aid for Scientific Research
under the grant numbers,
    13135201,		
    15540243,		
and 1707005.		


\section*{References}
\begin{thebibliography}{10}
\bibitem{Kawamoto}
Kawamoto N and Smit J
	1981\ {\it Nucl.\ Phys.}\ B\ {\bf 192}\ 100;
Damgaard P H, Kawamoto N and Shigemoto K
	1984\ {\it Phys.\ Rev.\ Lett.}\ {\bf 53}\ 2211
\bibitem{Faldt1986}
Faldt G and Petersson B
	1986\ {\it Nucl.\ Phys.}\ B\ {\bf 265}\ 197
\bibitem{Bilic1992}
Bili{\'c} N, Karsch F and Redlich K
	1992\ {\it Phys.\ Rev.}\ D\ {\bf 45}\ 3228
\bibitem{Nishida2004}
Nishida Y
	2004\ {\it Phys.\ Rev.}\ D\ {\bf 69}\ 094501
	[arXiv:hep-ph/0312371]
\bibitem{Tsubakihara}
Tsubakihara K and Ohnishi A
	2006 
	arXiv:nucl-th/0607046
\bibitem{KMOO_2007}
Kawamoto N, Miura K, Ohnishi A and Ohnuma T
	2007\ {\it Phys.\ Rev.}\ D\ {\bf 75}\ 014502 
	[arXiv:hep-lat/0512023]
\bibitem{Azcoiti2003}
	Azcoiti V et al.
	2003\ {\it J. High Energy Phys.} {\bf 0309}\ 014\ 
	[arXiv:hep-lat/0307019]
\bibitem{Bilic1995}
Bili{\'c} N and Cleymans J
	1995\ {\it Phys.\ Lett.}\ B\ {\bf 355} 266\ 
	[arXiv:hep-lat/9501019]
\bibitem{Ipp}
	Ipp A, in this proceedings;
	Ipp A et al.
	2006\ {\it Phys.\ Rev.}\ D\ {\bf 74} 045016\ 
	[arXiv:hep-ph/0604060]
\end{thebibliography}
\end{document}